# Flow induced by a sphere settling in an aging yield-stress fluid.

B. Gueslin,[a), b)] L. Talini,[a)c)] B. Herzhaft,[b)] Y. Peysson,[b)] and C. Allain[a)]

[a)] Université Pierre et Marie Curie-Paris6, UMR 7608, Orsay, F-91405 France ; CNRS , UMR 7608, Orsay, F-91405 France ; Université Paris-Sud 11, UMR 7608, Orsay, F-91405 France.

[b)] Institut Français du Pétrole, 1 et 4 avenue de Bois Préau, 92852 Rueil Malmaison, France

[c)] corresponding author: Laboratoire FAST, Bât. 502, Campus Universitaire, 91405 Orsay Cedex, France, talini@fast.u-psud.fr

**Abstract:** We have studied the flow induced by a macroscopic spherical particle settling in a Laponite suspension that exhibits a yield-stress, thixotropy and shear-thinning. We show that the fluid thixotropy (or aging) induces an increase with time of both the apparent yield stress and shear-thinning properties but also a breaking of the flow fore-aft symmetry predicted in Hershel-Bulkley fluids (yield-stress, shear-thinning fluids with no thixotropy). We have also varied the stress exerted by the particles on the fluid by using particles of different densities. Although the stresses exerted by the particles are of the same order of magnitude, the velocity field presents utterly different features: whereas the flow around the lighter particle shows a confinement similar to the one observed in shear-thinning fluids, the wake of the heavier particle is characterized by an upward motion of the fluid ("negative wake"), whatever the fluid's age. We compare the features of this negative wake to the one observed in viscoelastic shear-thinning fluids (polymeric or micelle solutions). Although the flows around the two particles strongly differ, their settling behaviors display no apparent difference which constitutes an intriguing result and evidences the complexity of the dependence of the drag factor on flow field.

# I. INTRODUCTION

In the past years, the settling of a spherical particle through a non-Newtonian medium has been the object of numerous studies either numerical or experimental.[1,2] Owing to the various rheological properties of non-Newtonian fluids, the resulting fluid flows exhibit a broad range of complex and unexpected features, some of them remaining poorly understood although the problem of an isolated settling sphere is the simplest that can be encountered in sedimentation. In addition to the fundamental interest of the subject, a better understanding of the involved phenomena is also required for various industrial applications such as drilling of oil and gas wells.

In a Newtonian fluid, the creeping flow around a settling sphere presents a fore-aft symmetry. This symmetry is broken in non-Newtonian fluids exhibiting various rheological properties even at small Reynolds numbers.[3-5] For instance a negative wake i.e. an upward motion of the fluid in the particle's wake can be observed in polymeric fluids that are viscoelastic and shear-thinning.[4,6] Although this effect is well known, its underlying physical mechanism is still in debate; some authors[7] have attributed it to an interplay between shearing stresses in the particle's wake that drive a flow away from the sphere and extensional stresses that drive a flow toward the sphere, whereas others[8] invoke a competition between extensional stress and normal stress differences effects.

In the case of yield-stress fluids, numerical results have shown that during the creeping motion of a sphere in an unbounded Bingham fluid (i.e. a yield-stress fluid with a constant viscosity and no thixotropy), the flow is confined in the vicinity of the sphere within an envelope of fluid whose size depends on the value of the yield stress.[9] Outside of this envelope the stress is smaller than the yield stress and the fluid is motionless. Similar numerical results have been obtained in tubes filled either with Bingham fluids[10] or with Herschel-Bulkley fluids that exhibit both yield stress and shear-thinning.[11] The shape of the yielded region indeed depends on both the sphere to tube diameter ratio and the variations of the viscosity with the shear rate, but in any case the flow exhibits fore-aft symmetry, and the size of the yielded region is a decreasing function of the yield stress. These predictions compare well with measurements performed in polymeric yield-stress fluids that present both yield stress and shear-thinning.[12,13] Experimental data on the velocity field induced by the settling in a yield-stress fluid remains scarce however, owing in particular to the small number of yield-stress fluids that are suited for such experimental characterizations.

Among the existing experimental studies, only a few deal with the yield-stress fluids formed by aqueous clay suspensions. They yet constitute systems of interest for sedimentation studies since, from a fundamental point of view, their structure at the microscopic scale (and thus rheological behavior) utterly differs from the one of the more studied yield stress polymeric fluids and, from a practical point of view, they are involved in many industrial processes. Such fluids have aroused interest in the past years due to their aging properties. Owing to their microscopic structure, which is still a matter of debate,[14-17] the state of these fluids evolves with time in an irreversible way that depends on their mechanical history. In particular, when at rest, their apparent yield stress increases with time.[18] From a rheological point of view, they constitute thixotropic fluids that are also shear-thinning above a yield stress. It has been shown that in such fluids the settling behavior of a spherical particle is more complex than in simple yield-stress fluids[19]. Whereas in the latter fluids one expects the particle velocity to be either zero or of a constant value according to the relative values of the yield stress and the stress exerted by the particle, in Laponite suspensions a settling regime characterized by a decreasing particle velocity has been observed. This behavior results from the aging of the fluid as the particle settles.

The aim of the present work is to characterize the creeping flow induced by a spherical particle settling in a yield-stress fluid that exhibits aging properties, a Laponite suspension.

We show in particular that the characteristics of the velocity field can be strongly modified according to both the stress exerted by the particle on the fluid and the fluid's age. The paper is organized as follows: in section II we describe the materials and experimental procedures, whereas section III is devoted to the experimental results, and section IV to the main conclusions.

## II. MATERIALS AND PROCEDURES

### A. Experimental set up

The settling cell is made of Plexiglas, is of rectangular section (inner dimensions 12 cm x 2.4 cm) and of useful height 8 cm. Once filled with the fluid it is sealed with a cap that has been specially designed to release several particles in the center-plane of the cell parallel to its largest walls. The cap consists of a small tank whose bottom is drilled at regular intervals along the length of the cell; the particles are immersed in the tank and further released through the holes using tweezers. This set-up enables us to probe the same sample of fluid at different aging times, which is essential considering the poor reproducibility of Laponite suspensions. Only one particle is released through each hole, and we have verified that the distance between two neighboring holes is large enough to ensure that the behavior of one particle is not influenced by the particles released through other holes.

The fluid velocity is measured using the Particle Image Velocimetry (PIV) technique. The fluid is therefore seeded using glass hollow micro-spheres (diameter 10 µm) of density close to that of the fluid. The concentration of micro-spheres (0.01% in weight) is small and the fluid rheology is not modified within the time scale of the experiments. The settling plane is illuminated by a laser sheet produced from a He-Ne laser. The width of the sheet is 0.5 mm. The motions of the micro-spheres are imaged using a CCD camera and further digitized. The velocity field is computed using a PIV commercial software (Davis, LaVision) that cross-correlates two successive images separated by a time interval $\Delta t$. A typical image is shown in Fig. 1, where the horizontal and vertical axes are defined. The optimal range of $\Delta t$ has been found to lie between the time needed by the macroscopic particle to move over its radius and twice that time (the latter corresponding to a displacement of 60 pixels with the chosen magnification). The horizontal and vertical components (respectively $v_x$ and $v_z$) of the fluid velocity in the laboratory frame are then obtained from the PIV computations. Inherent sources of error upon the flow velocity originate from the finite size of the micro-spheres, as well as their "twinkling" in the laser light. Tests using reference motions have shown that the resulting relative error upon the velocity is approximately 5%. Note also that the reflection of the laser light on the particle and the particle's shadow (as can be seen in Fig. 1) hinder the micro-spheres' motion respectively on the right and left sides of the particle. The velocity along the horizontal axis of the particle can therefore be determined only at sufficiently large distances from the particle.

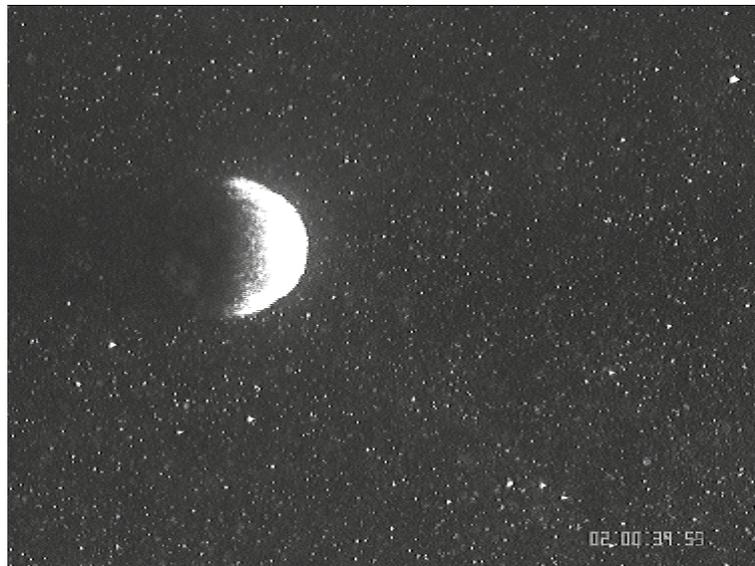

*Fig. 1: Typical image for PIV measurements. The particle (diameter 1.5 mm) is moving to the bottom of the picture. The fluid that is seeded with glass microspheres is illuminated from the right of the picture using a laser sheet.*

Calibrated macroscopic spherical particles of radius $a = 0.75$ mm and of two different densities have been used: aluminum particles (density 2.7 g/cm$^3$) and brass particles (8.7 g/cm$^3$). The velocity $V_p$ of the macroscopic particles is measured at different aging times using an image processing software (Image Pro +, BFIOptilas) with a precision of 15%. In the case of the PIV measurements, the values of $V_p$ used to normalize the fluid velocity have been measured directly from the PIV images with a precision of 5%. As shown in the following, the velocity of the particles strongly varies according to the fluid's age. The characteristic stress the particles exert on the fluid, $\sigma_p$, can be simply estimated as the ratio of the buoyancy forces and the sphere's section which yields $\sigma_p = 4/3 \Delta\rho g a$ where $\Delta\rho$ is the difference of densities of the fluid and particle and $g$ the acceleration of gravity.

A more realistic approach consists in taking into account the flow induced by the settling sphere: we thus assume that the characteristic shear rate $\dot{\gamma}_p$ scales as $V_p/2a$. Further using the relation $\sigma_p = \eta \dot{\gamma}_p$ and equating the viscous friction with the gravitational and buoyancy forces, finally yields:

$$\sigma_p = \frac{\Delta\rho g a}{9} \qquad (1)$$

The values of $\sigma_p$ given by equation (1) are respectively 1.4 Pa for the aluminum particles and 6.3 Pa for brass particles.

**B. Fluid**

The fluid is an aqueous suspension of Laponite (Rockwood) that is well known to constitute a transparent thixotropic fluid. Since the physico-chemical conditions can dramatically modify the state of the suspensions, special care has been paid to their preparation: sodium hydroxide is added to pure water so as to obtain a solution of pH 10; the ionic strength is therefore of $10^{-4}$ M. Laponite is further added at a concentration of 2.5% w/w under a nitrogen atmosphere, and the suspension is strongly mixed. The samples are then kept under a nitrogen atmosphere for ten days during which they are regularly stirred to ensure a good hydration of the clay particles. Prior to the settling experiments the suspension is stirred at a velocity of 1000 rpm for 1h in order to obtain a reproducible initial state. The fluid is further poured in the settling cell and left at rest for a time we denote as the aging time $t_a$. The investigated aging times range from 30 to 220 minutes, during which the fluid evolves from an apparently fluid state to a pasty material.

The rheological properties of the Laponite suspensions have been the object of several studies in the past years. It has thus been shown that Laponite suspensions are thixotropic yield-stress fluids that present shear-thinning properties above the yield stress.[20] The accurate determination of the rheological properties of the Laponite suspensions is nevertheless not straightforward. In particular, owing to the aging of the fluid, the yield stress is not an intrinsic property but depends on the mechanical history of the fluid.[21] Henceforward, although it is currently stated in the literature that the yield stress of Laponite suspensions increases with age, the nature of this variation has not been systematically documented. As far as we know the only attempt to measure this variation has been performed with local rheological probes; the authors found that the yield stress increases with the fluid aging time and may saturate at large aging times.[22]

In order to characterize the fluid rheology, we have performed stress ramp tests using a controlled stress rheometer. The fluid is first strongly stirred as for a settling experiment and poured in a cylindrical, double gap Couette flow apparatus. At a given aging time the fluid is submitted to a shear stress, increasing at a constant ramp, and the response in shear rate is measured. Although no steady-state is reached, these tests provide characterization of the fluid rheology with a weak contribution of the aging since the duration of a measurement (ten minutes) is smaller than the aging times investigated. As shown in the stress vs. shear rate plots of Fig. 2, the behavior is similar for all aging times: at large shear stresses the fluid is shear-thinning and the shear stress and shear rate are related through a power law equation:

$$\sigma = m\dot{\gamma}^n \qquad (2)$$

The parameter $m$ increases for increasing aging times while the index $n$ decreases, indicating that the more the fluid ages the more it shear thins. At small shear stresses, although equation (2) does not hold, the behavior is still shear-thinning and no yield stress value can be inferred from the rheological measurements. These results are not surprising considering the difficulties evoked above in characterizing the yield stress of Laponite suspensions. In our case, the value of the yield stress being small, more adapted apparatus and procedure would be required for an accurate measurement. We have not attempted to further characterize the fluid rheology since it is beyond the scope of this paper.

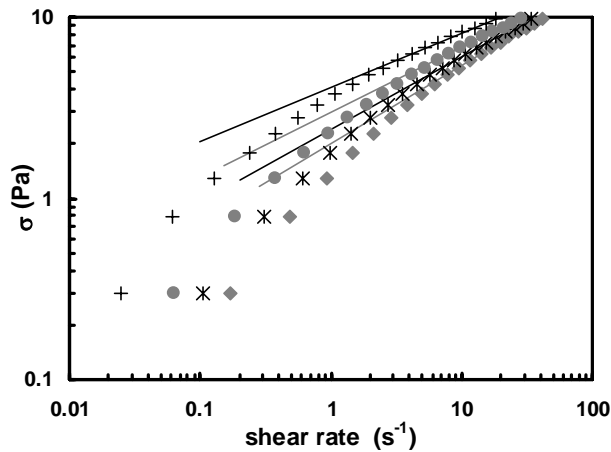

*Fig. 2: Rheograms obtained with a stress controlled rheometer for different values of the aging time: increasing values of the stress are applied to the fluid and the shear rate is measured for each value of the shear stress. At large shear stresses the shear stress and shear rate are related through the power-law equation (2) with $m = 2.0$ S.I. and $n = 0.42$ for $t_a = 45$ min (diamonds), $m = 2.5$ S.I. and $n = 0.39$ for $t_a = 60$ min (stars), $m = 3.0$ S.I. and $n = 0.35$ for $t_a = 90$ min (dots), $m = 4.1$ S.I. and $n = 0.30$ for $t_a = 120$ min (crosses).*

## C. Particle velocities in the Laponite suspension

As shown in Fig. 3, the measurements of the sphere settling velocities reflect the fluid temporal evolution: the velocities vary by two orders of magnitude within the investigated aging time scale. Note that as previously observed,[19] at small aging times the measured velocity is large such that its variation along the height of the settling cell remains very small, whereas at large aging times the motion is much slower and the velocity significantly decreases.

The variations of the velocity of each particle are well described by:

$$V_p = V_0 e^{-t_a/\tau} \qquad (3)$$

where $V_0$ is the settling velocity at $t_a = 0$ and $\tau$ a time constant that represents a characteristic aging time. Both parameters depend on the density of the particle: we have found $V_O = 6.0 \pm 1.2\, mm/s$ and $\tau = 16.5 \pm 1.7\, min$ for the lighter (aluminum) particle, and $V_O = 50 \pm 10\, mm/s$ and $\tau = 23.5 \pm 2.4\, min$ for the heavier (brass) particle.

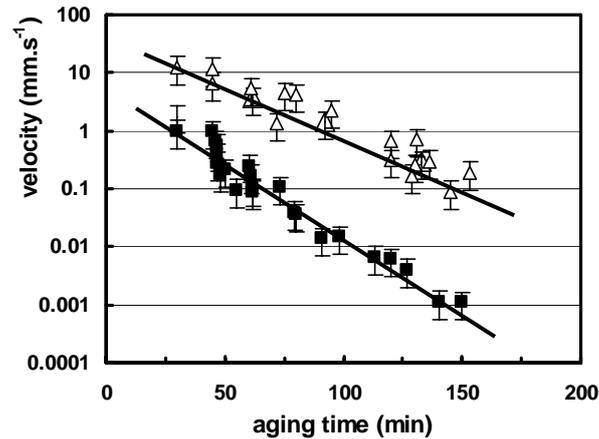

*Fig. 3: Settling velocities of the aluminum (full squares) and brass (triangles) particles in the Laponite suspension as a function of its aging time. The experimental data is well described using equation (3) (full lines).*

These values are not consistent with the simple dimensional argument previously given to estimate the stress exerted by the particles on the fluid. Following this argument, one would expect that $\frac{V}{2a}\left(\frac{m}{\sigma_p}\right)^{1/n} = cste$; Fig. 4 shows that this quantity is not constant but instead decreases by an order of magnitude within the investigated range of aging times. This decrease may be attributed to a dependency of $\frac{V}{2a}\left(\frac{m}{\sigma_p}\right)^{1/n}$ on the shear-thinning index as has been observed in shear-thinning fluids,[1] or to an increasing contribution of the yield stress to the total stress as the aging time increases.

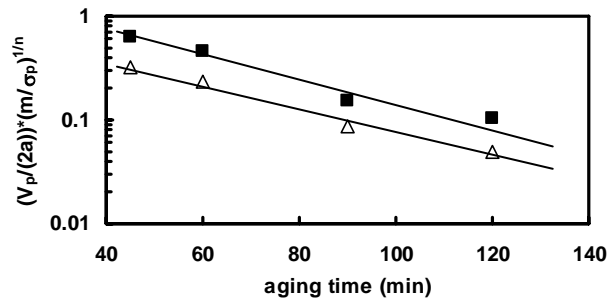

*Fig. 4: Variations with the fluid aging time of the dimensionless characteristic shear rate of the aluminum (full squares) and brass (triangles) particles. The settling velocity $V_p$ is directly measured whereas $\sigma_p$ is given by equation (1) and parameters m and n are deduced from the rheological measurements. The full lines are guides to the eye.*

Information on the yield properties of the fluid is provided by the settling experiments: at large enough aging times the particles are observed to stop and remain suspended at a given height of the cell for times as long as several days. Following equation (3), the particle reaches its equilibrium position at infinite aging times. We therefore characterized the stoppage of the particles by measuring the vertical position of the particles that were apparently motionless for more than ten hours. The measurements are in agreement with the equilibrium positions deduced from equation (3) within the uncertainties of the parameters $V_0$ and $\tau$. We can therefore predict the aging time at which one particle reaches the position 5 μm above its equilibrium position (5 μm being the experimental uncertainty upon the position).

For aluminum particles ($\sigma_p = 1.4\,\text{Pa}$) we have found $t_a = 230 \pm 30\,\text{min}$ and $t_a = 390 \pm 50\,\text{min}$ for brass particles ($\sigma_p = 6.3\,\text{Pa}$). The increase of the yield stress with the aging time is in agreement with results obtained in a Laponite suspension of similar concentration (2.7% in weight).[22] The discrepancy in the estimated values may result from the different geometry and length scale probed in the given reference, where a dynamic yield stress is deduced from the rotating motion of a magnetic needle under the effect of an applied magnetic torque. Differences in the preparation of the suspensions may also explain this disagreement; in particular, in contrast to the cited work, our fluids are not filtered, which has a probable consequence on the microscopic structure of the suspensions.[15]

The particle Reynolds number can be estimated as classically in shear-thinning fluids:

$$Re = \frac{2^{n-1} \rho_f V_p^{2-n} a^n}{m} \quad (4)$$

where $\rho_f$ is the fluid density.

Using equation (3) together with the experimental values of the particle, the maximum Reynolds number is of the order of $10^{-2}$. The effects of inertia are therefore expected to be small in all the experimental results we present.

**D. Experiments in a Newtonian fluid**

Preliminary experiments have been conducted in a Newtonian fluid (pure glycerol) in order to calibrate the PIV measurements. Aluminum particles were used and the Reynolds number was $Re = 2 x 10^{-3}$. Fig. 5 shows the corresponding velocity field around the settling sphere. It exhibits a fore-aft symmetry as expected at small $Re$. Note that this symmetry is not obvious in Fig. 5 owing to the representation of the PIV vectors: each vector corresponds to the velocity of the point lying at its origin. The indicated position of the sphere corresponds to its average position between times $t$ and $t + \Delta t$.

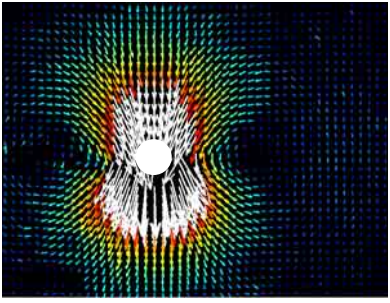

*Fig 5: Velocity field in the Newtonian fluid. The white circle indicates the average vertical position of the particle.*

The recirculating regions on both sides of the sphere are consequences of the finite size of the cell. The centers of these regions are located at (9.3 ± 0.2) radius from the sphere's center which roughly corresponds to mid-distance from the cell sidewall and the sphere's center. This result is in agreement with experimental data obtained in a cylindrical cell.[4] To our knowledge no data on the settling in a rectangular cell is available in the literature. Nevertheless, within the distances we probe, the lack of axisymmetry of the cell does not significantly modify the velocity field from the one in a cylindrical cell.

Fig. 6 shows the variations of the dimensionless z-component of the fluid velocity as a function of the vertical position along the sphere's centerline. The experimental data is compared with the Stokes' calculations for the velocity in an unbounded fluid. Away from the sphere the velocity decreases to zero faster than in an unbounded fluid and significant differences are observed for $z/a \geq 5$. The downstream and upstream velocity disturbances nevertheless extend farther than the visualization window. Note that in a cylindrical cell they are expected to be confined within a distance close to the width of the cell.[4]

Close to the sphere's surface ($z/a \leq 2$) large discrepancies from Stokes' calculations are observed (experimental data not shown) owing to the principle of PIV computations: the vectors obtained by PIV do not correspond to the instantaneous velocities of fluid particles, but to their displacements over a time spacing Δ*t* during which the sphere itself moves (*z* = 0 corresponds to its average vertical position between times $t$ and $t+\Delta t$). The correction resulting from this temporal averaging can be computed by integrating the expression for the fluid velocity over $\Delta t$, and we have found it to be in agreement with the experimental data; no other source of error is therefore involved in the observed discrepancy. The correction is close to 20% for $z/a =1.5$, whereas it becomes smaller than the experimental uncertainty for $z/a=2.5$.

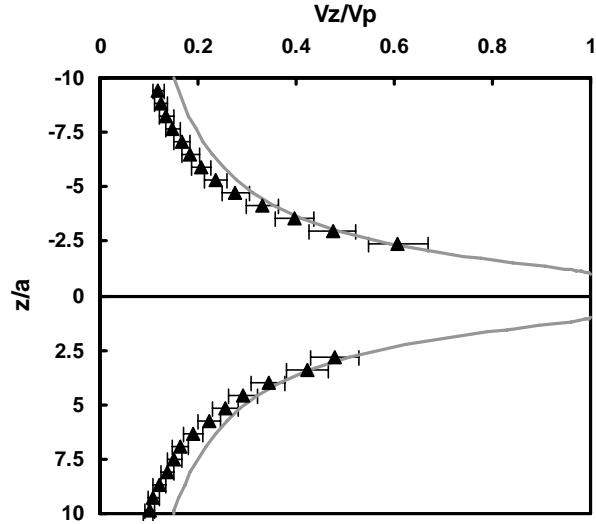

*Fig. 6: Vertical component of the fluid velocity along the vertical centerline of the aluminum sphere in the Newtonian fluid. The full line represents the Stokes' solution.*

In conclusion, despite their unreliability in the vicinity of the sphere's surface (within 1 radius), the PIV measurements show a good agreement with the velocity field expected in a Newtonian fluid.

**III. VELOCITY FIELDS IN THE LAPONITE SUSPENSION**

**A. Qualitative description**

Figure 7 shows the velocity fields around the aluminum and brass particles at the same aging time. Note that the vectors scales are different in both figures since the particle velocities differ by two orders of magnitude. The fore-aft symmetry observed in a Newtonian fluid is broken for both particles in the Laponite suspension. In both cases the flow therefore qualitatively differs from the one predicted and observed in non thixotropic yield-stress fluids.[9-12] Note that, in the cited experimental study[12], the sphere being moved at a constant speed for the flow visualization, a rod is attached at its rear stagnation point, which may make it difficult to observe fore-aft dissymmetry of the flow (although the authors evoke a complete fore and aft symmetry).

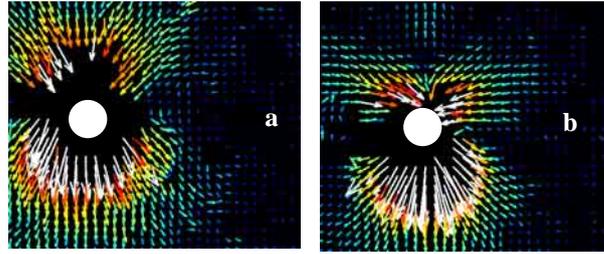

*Fig. 7: Velocity fields in the Laponite suspension for (a) the aluminum and (b) the brass particles. Both velocity fields were obtained at an aging time $t_a = 80$ min.*

Although the symmetry is broken in both cases, the downstream velocity fields are nevertheless utterly different according to the particle's weight: whereas the velocity decreases monotonically away from the lighter (aluminum) particle, a "negative wake", i.e. an upward flow in the particle's wake, is observed downstream of the heavier (brass) particle whatever the aging time. The existence of a negative wake has been repeatedly reported in polymeric fluids[4,6] or in wormlike micelle solutions,[23] but had so far never been observed in a yield-stress fluid to our knowledge. To complete this qualitative picture of the velocity fields, let us note that although they do not appear clearly in Fig. 7, recirculating regions have been observed on both sides of the spheres. This point is detailed in the following paragraph.

**B. Recirculating regions**

Similar recirculating regions as in the Newtonian case are observed on each side of the particle. We have determined the horizontal position $x_s$ of the centers of these recirculating regions as a function of the aging time for both aluminum and brass particles. The results that are reported in Fig. 8 show that in both cases the confinement of the flow in the horizontal direction increases with the aging time, i.e. the centers of the recirculating regions move closer to the sphere. The corresponding decrease of $x_s$ is stronger for the brass than the aluminum particle. Note that we have checked that no wall slip occurs at the surface of the sphere, as has been observed in other yield stress fluids.[24]

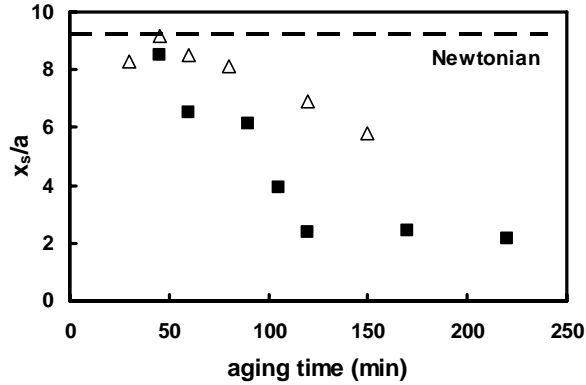

*Fig. 8: Horizontal position $x_s$ of the centers of the recirculating regions on the sides of the aluminum (full squares) and brass (triangles) particles as a function of the aging time. The dashed line indicates the position measured in the Newtonian fluid and the hatched area the domain in which measurements are possible.*

The decrease of $x_s$ with the aging time is qualitatively similar to results obtained in simple yield-stress fluids for which the size of the sheared zone moving with the particle decreases with the increasing yield stress.[9-12] Let us note that one would expect the flow to be more confined for the aluminum than the brass particle ($\sigma_p$ being smaller and therefore closer to the yield stress for the aluminum particle) which is not the case. Since the size of the sheared zone in particular depends on the values of the yield stress and shear thinning, a more quantitative comparison of our results with the numerical simulations would require a better knowledge of the fluid rheology.

**C. Upstream and downstream regions**

We now focus on the upstream and downstream velocities. More quantitative information is displayed in Fig. 9 and 10 that show for both spheres the variations of the dimensionless z-component of the fluid velocity as a function of the vertical position along the sphere's centerline and for different aging times.

Upstream of the lighter (aluminum) particle, the velocities are below the Newtonian limit. This slight deviation has been observed in polymeric shear-thinning fluids, and is attributed to a decrease in the fluid viscosity as the shear rate increases closer to the sphere[4]. As shown in Fig. 9, no significant influence of the aging time on this deviation is observed. This is not the case downstream of the particle, where the velocity is much smaller than in a Newtonian fluid and decreases faster away from the particle as the aging time increases. As mentioned earlier, the shear-thinning properties of the Laponite suspension increase with the aging time, which is consistent with the increasing confinement of the downstream flow.

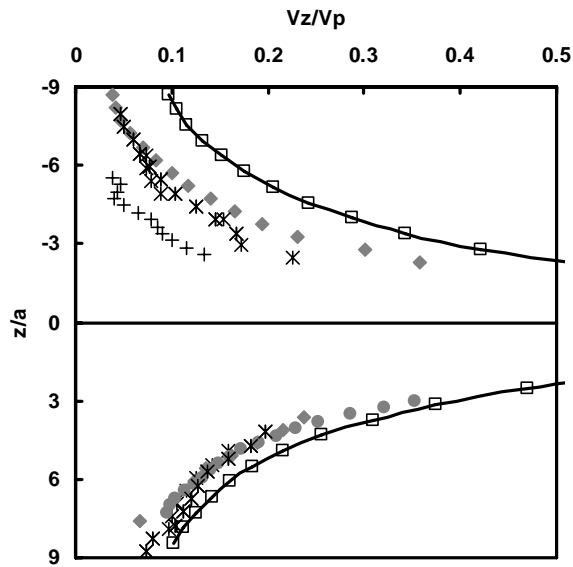

*Fig. 9: Vertical component of the fluid velocity along the vertical centerline of the aluminum sphere in the Laponite suspension at $t_a = 45$ min (diamonds), $t_a = 60$ min (stars) and $t_a = 120$ min (crosses). The full line and squares correspond to the fluid velocity around the same particle measured in the Newtonian fluid.*

In the case of the heavier (brass) particle, the upstream flow is even more confined and, as observed in the horizontal direction (results not shown), its confinement increases with the aging time which is once more consistent with the existence of yielded and unyielded zones. The presence of a negative wake, however, prevents further comparison between this flow and the one in a simple yield stress fluid.

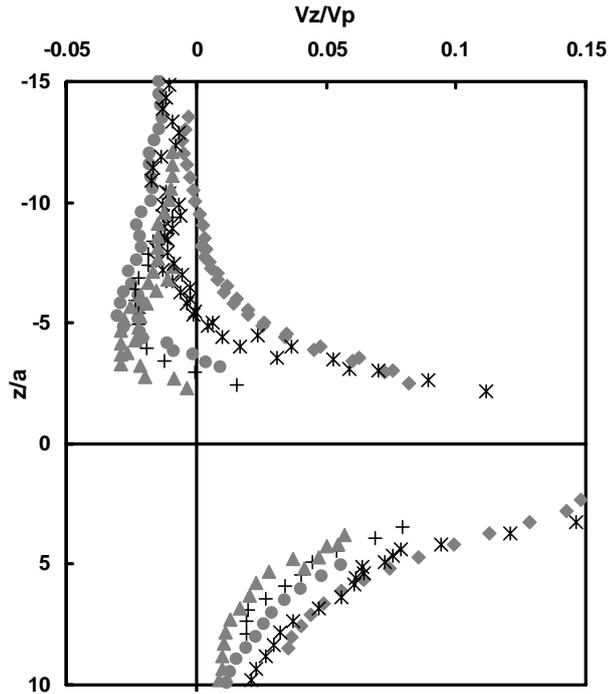

*Fig. 10: Vertical component of the fluid velocity along the vertical centerline of the brass sphere in the Laponite suspension at $t_a = 45$ min (diamonds), $t_a = 60$ min (stars) and $t_a = 105$ min (full circles), $t_a = 120$ min (crosses) and $t_a = 170$ min (full triangles).*

The features of the negative wakes shown in Fig. 10 are furthermore quantitatively different from the ones observed in shear-thinning polymeric fluids. In particular, the stagnation point in the particle's wake moves closer to the particle as the aging time increases. In polymeric shear-thinning fluids, the position of the stagnation point has been found to be independent of the Deborah number i.e. the ratio of the characteristic fluid relaxation time with the flow characteristic time, the latter scaling as $a/V_p$.[4] In micelle solutions however the stagnation point moves away from the sphere as *De* increases.[23] In an aging fluid such as a Laponite suspension, the only characteristic time is the aging time. Taking into account the variations of the velocity with the aging time as described in equation (3), the Deborah number is thus expected to vary as $t_a e^{-t_a/\tau}$. Since the experiments are performed within $t_a > \tau$ for both particles, one expects the Deborah number to decrease with increasing aging times. Our results are therefore in qualitative agreement with the ones obtained in micelle solutions.

**IV. CONCLUSIONS**

We have shown that the features of the flow induced by a particle settling in an aging yield-stress fluid strongly depend on the characteristic stress exerted by the particle. Despite the stresses exerted by the two chosen particles being of the same order of magnitude, the flow confinement as well as the velocity in the particle's wake strongly differs from one another. We have also shown that no symmetric sheared zone around the particle is observed as has been predicted and observed in non thixotropic, yield-stress fluids. Thixotropy therefore not only induces time-dependent rheological properties, but also results in a breaking of the fore-aft symmetry. Whereas the existence of a yield stress can be inferred from the settling behavior, its influence on the velocity field around the particles is not obvious.

Although the exact criterion for the presence of a negative wake is still in debate[2], the fluid extensional properties doubtlessly play a crucial role in the apparition of a negative wake. A better knowledge of the extensional rheology of the Laponite suspensions is therefore needed to understand the present results. Such measurements have never been performed as far as we know. Laponite suspensions being constituted of colloidal disks whose interactions remain weak, their extensional viscosity is expected to be small compared to the one of polymeric fluids; accurate measurements of the extensional viscosity of Laponite suspensions should therefore constitute a difficult task.

Finally let us note that although the flows around the two different particles are qualitatively different, the settling behaviors remain similar. The correlation between the velocity field and the drag factor is therefore not obvious in the case of the Laponite suspensions.

Acknowledgements : We thank F. Moisy for his help in the PIV experiments. The thesis of B. Gueslin is partially funded by ANRT.